# The Abundance of Deuterium and He3 in the Solar Wind

Frank Scherb

Physics Dept., University of Wisconsin, Madison, WI 53706
scherb@physics.wisc.edu

## Abstract

The relative abundance of deuterium (D) in the solar atmosphere is not known. D is not only destroyed in stars, it is also synthesized in the atmospheres of active stars (Prodanovic and Fields 2003). In several cases, production of D in the sun has been detected when solar flares occur, using both energetic particle measurements (Anglin 1975) and by detection of 2.223 MeV gamma rays emitted by D (Terekhov et al. 1996; Shih et al. 2009). We describe a project to measure the abundance of deuterium in the solar wind, and to monitor its evolution during a several-year period. The instrument consists of two grids, a tritium target, and semiconductor particle detectors. The grids, which are hemispherical and concentric, accelerate the incident solar wind ions using a potential difference on the order of ~80 to 100 kV and concentrate the ions on the tritium target. A fraction of the solar wind deuterons thus accelerated interact with the target to produce 3.6 MeV alpha particles, some of which are recorded by adjacent semiconductor detectors. A similar instrument was successfully tested in space in 1975 in order to observe positive auroral ions in a hydrogen aurora.

## The Experiment

While in situ space measurements have shown the possibility of detecting high- energy (~a few MeV) deuterons linked to solar flares (Anglin 1975), the same is not true for less energetic solar wind deuterons (~< 10 keV). Therefore, we describe a project that uses a new method for detecting solar wind deuterium. This method is based on the nuclear reaction between D and tritium:

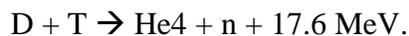

D + T → He4 + n + 17.6 MeV.

The neutron and alpha particle that are produced have energies of 14.0 and 3.6 MeV respectively. **Figure 1** shows that the cross section of this reaction possesses a maximum for deuterons with energies around 100 keV. Since the energy of deuterons in the solar wind is below 10 keV, an instrument designed to use the reaction D + T must accelerate the deuterons to energies greater than 50 keV, in order to attain appreciable values for the cross section (Floyd and Scherb 1966).  To accelerate the D particles of the solar wind, we generate an electric field between two hemispherical grids (Reynolds and Scherb 1967) by applying a potential difference on the order of 100 kV.  The high- voltage accelerator was successfully tested in space on a high-altitude Javelin sounding rocket in 1975 in order to measure energetic auroral ions in the upper atmosphere of the Earth ( Lynch et al. 1976; Lynch et al. 1977).

He3 ions in the solar wind will also react with the tritium target. The He3 + T reaction has three possible final states:

(a) He4 + D +14.3 MeV           (43%)

(b) He4 + n + p +12.1 MeV        (51%)

(c) He4 + p + n + 14.1 MeV       (6%)

The He4 and D resulting from reaction (a) have unique energies of 4.75 and 9.55 MeV, respectively. Reactions (b) and (c) produce protons and alphas that do not have unique energies. Therefore, some of these particles will have energies near enough to 3.6 MeV to be confused with alphas from the D + T reaction.   **Figure 2** , however, shows that reaction (a) can be used to resolve the ambiguity, while also yielding information on the abundance of He3 ions in the solar wind.

Measurements carried out with a crude, early laboratory model of the experiment indicated that for a D/H ratio in the solar wind of $10^{-5}$, and an instrument entrance aperture of $10^{3}$ cm$^{2}$, the deuteron count rate would be about one count per sec in the slow solar wind (Floyd and Scherb 1966).

A campaign of observations spanning several years would not only produce a measure of the relative abundance of atmospheric D free of bias due to sporadic solar activity, but would also permit a better characterization of the production mechanisms of D in flares.

In **Figure 3**, a sketch of the instrument shows the solar wind particle accelerator, its position relative to the tritium target, and one of several alpha particle detectors.  Other particle detectors, optimized to make direct measurements of solar wind composition, would be integrated into the instrument.

References

Anglin, J. D. 1975, ApJ, 198,733

Floyd, F. W., & Scherb, F. 1966, IEEE Trans., NS-13, 18

Lynch, J., Leach, R., Pulliam, D., & Scherb, F. 1977, JGR, 82, 1951

Lynch, J., Pulliam, D., Leach, R., & Scherb, F. 1976, JGR, 81, 1264

Prodanovic, T., & Fields, B. D.  2003, ApJ, 597, 48

Reynolds, R., & Scherb, F. 1967, Rev. Sci. Instr., 38, 348

Shih, A. Y., Lin, R. P., & Smith, D. M. 2009, ApJ, 698, L152

Terekhov, O. V., Sunyaev, R. A., Tkachenko, A.Yu., Denisenko, D.V., Kuznetsov, A.V., Barat, C., Dezalay, J.-P., & Talon, R. 1996, Astron. Lett., 22, 143

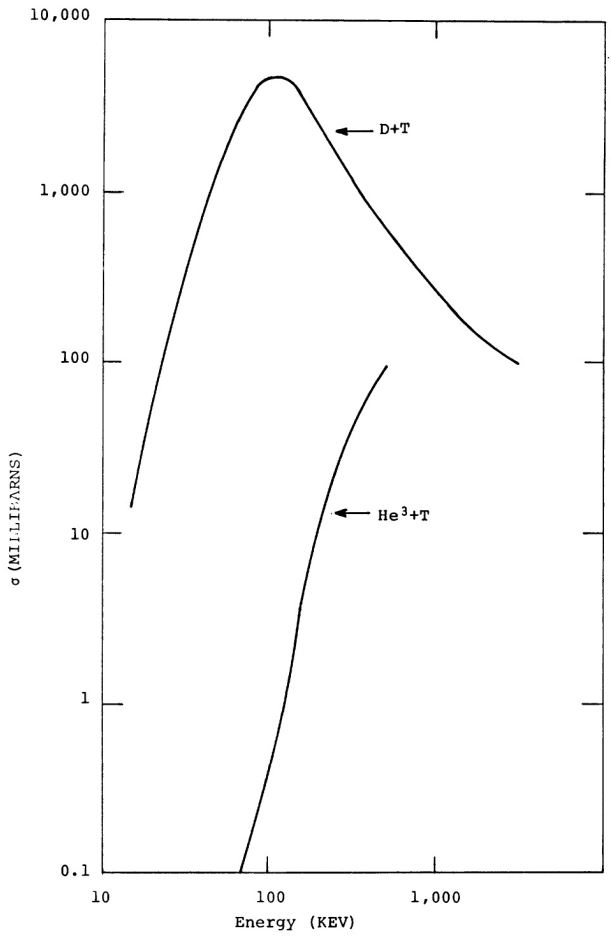

Fig. 1. Nuclear reaction cross sections

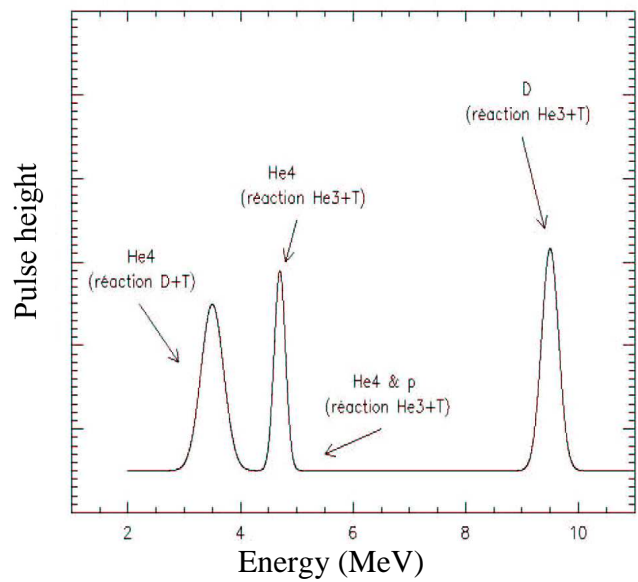

Fig. 2. Schematic pulse height spectrum of deuterium detector, showing effects of He3.

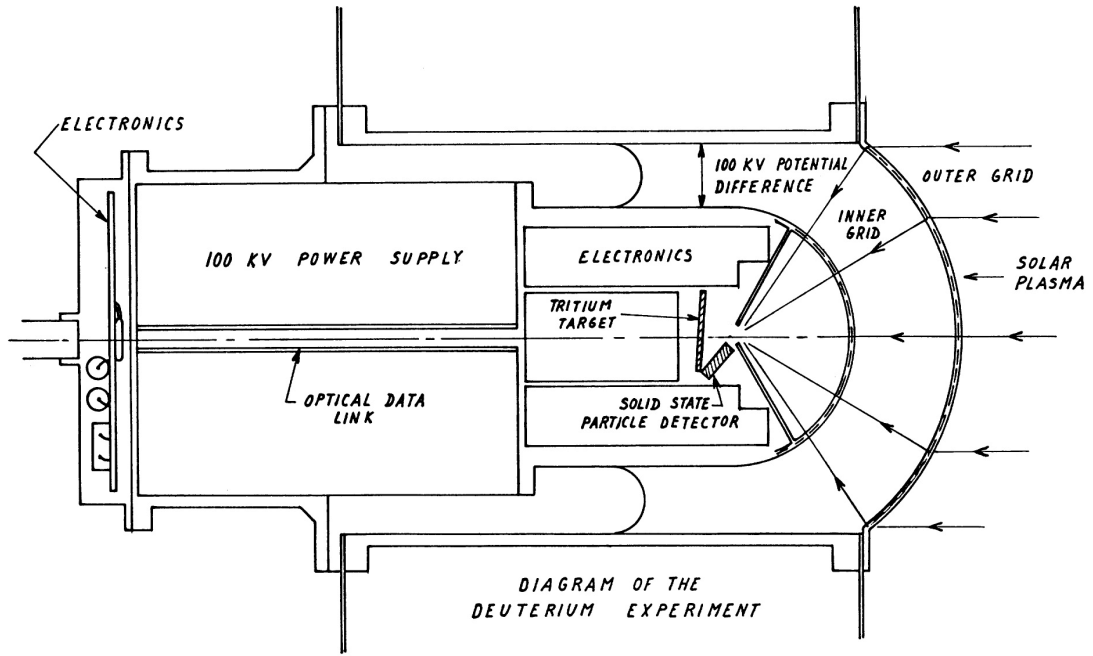

Fig. 3. Accelerator-detector instrument